\documentstyle[prl,aps,psfig]{revtex}
\newcommand{\be}{\begin{equation}}
\newcommand{\ee}{\end{equation}}

\begin{document}
\title{Drift and trapping in biased diffusion on disordered lattices}
\author{Deepak Dhar$^1$ and Dietrich Stauffer$^2$}
\address{$^1$ Department of Theoretical Physics, Tata Institute of Fundamental
Research, Homi Bhabha Road, Bombay 400005, India; e-mail :
ddhar@theory.tifr.res.in \\
$^2$Institute for Theoretical Physics, Cologne University, 
D-50939 K\"oln, Germany; e-mail: stauffer@thp.uni-koeln.de}

\maketitle
\widetext
\begin{abstract}
We reexamine the theory of transition from drift to no-drift in biased
diffusion on percolation networks. We argue that for the bias field $B$
equal to the critical value $B_c$, the average velocity at large times $t$
decreases to zero as $1/\log(t)$.  For $B < B_c$, the time required to
reach the steady-state velocity diverges as $\exp({\rm const}/|B_c-B|)$. 
We propose an extrapolation form that describes the behavior of
average velocity as a function of time at intermediate time scales. This
form is found to have a very good agreement with the results of extensive
Monte Carlo simulations on a 3-dimensional site-percolation network and
moderate bias. 
\end{abstract}

\bigskip

Diffusion in disordered lattices and in particular at their percolation
threshold is an old subject \cite{gennes}. If there is a preferred direction of
motion due to some external imposed field, we have biased diffusion. A
small value of the external bias gives rise to a mean displacement in the
direction of field that increases linearly with time, and the mean
velocity tends to a constant. This asymptotic value of the mean velocity
is proportional to the field for small fields. It was pointed out by
B\"ottger and Bryksin \cite{bryksin} that in disordered media, because of 
trapping in
dead-end branches, the mean velocity would be a nonmonotonic function of
the bias. In \cite{barma,dhar}, it was argued that for random walkers with no mutual
interactions, the asymptotic mean velocity actually becomes zero for a
finite value of bias $B_c$, and that the mean displacement increases as a
sublinear power of time for $B>B_c$. These theoretical arguments were
supported by exact calulation of the mean drift velocity on a random
comb \cite{white}.  A similar argument was used to show the existence of 
drift to no
drift transition in the case of non-interacting particles when the bias
field direction is not constant, but depends on the random geometry 
\cite{bunde}.
It was argued in \cite{ramaswamy} that this sharp transition from drift to 
no drift
disappears if a repulsive interaction between diffusing particles is taken
into account. 

      However, clear numerical verification of these theoretical
predictions has not been possible so far.  Earlier numerical simulations
failed to see clear evidence of a sharp transition from drift to no drift
as a function of the bias \cite{pandey}. For large bias, log-periodic oscillations in time
mask the possible transition \cite{sornette}. There are large sample to sample
fluctuations, and it appears that times much greater than feasible are
needed to see the asymptotic time regime for $B$ near $B_c$. This
has motivated  our reexamination of  this question in this paper.
 
      In this paper, we argue that the time needed to see the asymptotic
behavior predicted in \cite{barma,dhar} diverges as $\exp({\rm const}/|B_c-B|)$, as $B$
approaches $B_c$ from below. This kind of sharp increase of relaxation
times, often encountered in systems with quenched disorder (the familiar
Vogel-Fulcher law in glasses), implies that in order to compare the
theoretical predictions to simulations results, one must take into account
corrections to the true asymptotic behavior. We argue that at $B=B_c$, the
mean velocity decreases as $1/\log t$ for large times $t$. We propose a
simple extrapolation form that incorporates this behavior, and agrees with
the expected asymptotic behavior of mean velocity for large times. We test
this form by comparing to results of Monte Carlo simulations on a 3-d
simple cubic lattice. We find excellent agreement between the two.

  We start with a precise definition of the problem, and of the method
used in computer simulations. We consider a site percolation problem on an
$L \times L \times L$ simple cubic lattice with periodic boundary
conditions with concentration $p=1/2$ of sites colored white, and the rest
red. This value of $p$ is much larger than the critical percolation
threshold $p_c= 0.31160$, and for this value of $p$, more that 97.8\% of
the white sites belong to the infinite white cluster. At time $t=0$, a
group
of $N$ random walkers are placed on randomly selected white sites of the
lattice.  At each (discrete) time step, each walker attempts to move to a
nearest neighbor site, and actually moves there if the site is white. The
nearest neighbor is chosen to be the neighbour along the positive
x-direction with bias probability $B$, and to any of the six neighbours
with probability $1-B$; 
thus $B + (1-B)/6  = (1+5B)/6$ is the probability to move into
the positive x direction, and $(1-B)/6$ is the one for the negative x direction.
As in the earlier simulations \cite{pandey}, the status of
lattice sites was stored in single bits; thus $L = 964$ could be chosen
for most of our simulations. The total displacement of a walker
is calculated by adding up all her single-step displacements. This is
then averaged over the $N$ walkers. In our simulations, $N$ varies from
80,000 for the shorter time runs, to only $384$ for the longest simulations. 

  For weak bias $B$, the situation is quite straightforward. The average
velocity of the walkers quickly settles down to a constant value which for
small bias increases linearly with bias. In fig 1, we have plotted the
limiting velocity as a function of $B$, for $B$ lying between 0.15 and 0.47.
For larger $B$, the velocity decreases very slowly with time. We can not
be
sure of the true asymptotic velocity. The plotted mean velocities are
determined by looking at the rms
displacements  for $ 0.9 \times 10^9 < t< 
2.1 \times 10^9$. Also plotted in the figure are the limiting values
obtained by using the extrapolation formula (see below) for $B \ge .4$.
For lower values of $B$, these match very well. But for $B=.45 $ and
above, clearly much longer times would be needed to reach the asymptotic
values. 

  In the presence of strong bias, the diffusive motion of walkers in the
disordered lattice is slowed down because often the walker gets into a
cul-de-sac region, and it takes a long time to get out of them. For a
dead-end of depth $\ell$ the trapping time varies as $\lambda^{\ell}$,
where $\lambda$ is the ratio of transition probabilities along and
against the field. In our case $\lambda = (1+5B)/(1-B)$. In addition, 
according to Refs.\cite{barma,dhar} the
density of traps of depth $\ell$ decreases exponentially with $\ell$, say
Prob$(\ell)$ varies as $\exp(-A\ell)$, where $A$ is a $p-$dependent
constant. The average trapping time $<\tau>$ at a backbone site is given
by \cite{barma,dhar}

\be
<\tau> \sim \sum_{\ell=0}^{\infty} {\rm Prob}(\ell) \lambda^{\ell}
\ee
For $B > B_c$, this summation diverges, and the asymptotic drift velocity ,
which varies inversely as the trapping time, is zero. It is easy to see
that the critical value $B_c$ is determined by the equation

\be
(1+5B_c)/(1-B_c)\exp(-A)=1
\ee

For $B < B_c$, the average trapping time is finite, and the asymptotic
drift velocity $v_{\infty}$ is finite, and decreases to zero as $B$ tends
to $B_c$ from below. 
For $B>B_c$, the walker moves a distance $R$ of order 1/Prob$(\ell)$ before it
meets its first encountered trap of depth $\ell$. The time $t$  to escape from 
it is of the order of $\lambda^\ell$. Thus from eq(2) we have \cite{dhar}
$$R \propto \exp(A\ell) = \exp(A \ln t/\ln \lambda) = \exp(\ln t \ln \lambda_c/
\ln \lambda) = t^{\ln \lambda_c/\ln \lambda}$$ 
Thus the average distance moved increases as $t^{1-x}$ where

\be
x= 1 - {\log[(1+5B_c)/(1-B_c)]\over \log[(1+5B)/(1-B)]}
\ee

If $p=p_c$, the critical threshold for percolation, then the constant $A$,
and hence using eq. (2), $B_c$ is zero. In this case, $x=1$, and the mean
displacement grows only as a power of $\log t$. This has been observed in
simulations \cite{stauffer}. 

Consider now the motion of a large number of walkers at $B=B_c$ at large
times $t$. Let us estimate the average trapping time felt by one walker in
this time. This is approximately given by Eq. (1), except that the
summation is cut off at a value $\ell_{max}$, which is determined by the
condition that $\lambda^{\ell_{max}}$ is order $t$. This is because in
this time the walker is unlikely to have encountered a deeper trap. As
each term of this summation is roughly equal \cite{reason}, this implies that
average trapping time up to time $t$ increases as $\ell_{max} \sim
\log(t)$. Using the fact that average velocity is just the inverse of the
average trapping time, we see that

\be
v(t) \sim  1/\log t , \; {\rm for} \; B= B_c.
\ee

For $B = B_c -\epsilon$, the summation in Eq. (1) is finite, but diverges
as $1/\epsilon$ for small $\epsilon$. This implies that $v(t)$ tends to a
constant proportional to $(B_c-B)$ for $B < B_c$ for large $t$. It varies
as $1/\log t$ for $B=B_c$ and large $t$, and varies as $t^{-x}$ for
$B>B_c$, with $x$ determined by Eq. (3). A simple extrapolation form which
incorporates all these behaviors is

\be
v = Kx/ [ (t/t_0)^x -1]
\ee
where $K$ is a constant, independent of $t$, but weakly dependent on the 
bias field $B$. If $x$ is positive, $v$ decreases as $t^{-x}$ for large
$t$. For $x=0$
it varies as $1/\log t$, and for $x < 0$, it tends to a finite limit as $t
\rightarrow \infty$. For $B=B_c -\epsilon$, with $\epsilon$ small,
initially, up to some relaxation time $T$, the velocity will decrease
roughly as $1/\log t$, but for $t > T$ it reaches the constant asymptotic
value proportional to $\epsilon$. Matching these two values of velocity
at $t=T$, we get

\be
T \sim \exp({\rm const}/\epsilon)
\ee
Thus we see that the relaxation time of the system for $B$ near $B_c$
diverges according to the Vogel-Fulcher form, as was claimed in the
introduction. 

Eq.(5) suggests a simple way to analyse the simulation data to find
$B_c$. We plot $1/v$ versus  $\log t$ for various values of the bias field
$B$. The plot is linear  right at the possible transition
point, shows a decreasing slope, eventually settling to a finite constant
value for smaller $B$, and shows slope increasing with time  
for larger $B$. This is shown in Fig. 2. We find that for
$B=0.53$, the graph is fairly linear. The equation of the best fit
straight line is

\be
1/v = 11.2 \; \log (t/30) 
\ee

Comparing with the extrapolation formula, we see that for $x \rightarrow
0$, the right hand side reduces to $K/\log(t/t_0)$. This fixes the
parameters $B_c \simeq 0.53$, and $t_0 \simeq 30$. We have shown by
continuous curves
theoretical fits to the data in Fig 2, for other values of $B$, using
Eq. (3) to determine $x$ for a given choice of $B$. The only unknown
parameter $K$ is expected to depend on $B$ weakly. We determine its value
by selecting the best fit to the data. The best fit value of $K$ is
found to be 0.0678 for $B$=0.40, which increases to 0.0893 for $B$=.53 and
0.0927 for $B$=0.60. We see that by a suitable choice of
$K$, the extrapolation formula (5) provides a very good fit to the data in
the entire range of data  $t > 10^3$. For  $t < 10^3$, there are
significant  corrections due to short term transients, not taken into
account in the extrapolation form (5).

To better test Eq. (4), we have run the simulations precisely for $B=B_c$
for much longer times. The results are shown in Fig 3.  We see that no
significant deviations from the
linear behavior are seen for $t$ up to $> 10^{10}$. Note the large fluctuations
for large time data, which is averaged over only a small  number of
walkers. For shorter times these results were confirmed by P. Grassberger
(private communication) using a very different algorithm. 

               We see in Fig. 2 that with two global parameters ( $B_c$
and $t_0$), and only one free parameter $K$ for each bias value, we can
fit nicely the data for $ t > 10^3$ for all the bias fields $0.4 < B <
0.6$ to the functional form Eq. (5), making the above conclusions based on
it more trustworthy. Thus our data are in good agreement with the
theoretically predicted functional form, and provide the first direct
observation in simulations of a sharp transition between drift and no
drift. We expect a similar behavior to be present in the case of
'topological bias' studied in \cite{bunde}.

We thank M. Barma, P. Grassberger, D. Sornette, K. W. Kehr, A. Bunde and R. B.
Pandey
for encouraging discussions, HLRZ J\"ulich for computer time on the Cray
T3E, and SFB 341 for support.

\bigskip

\begin{figure}
\begin{center}
\leavevmode
\psfig{figure=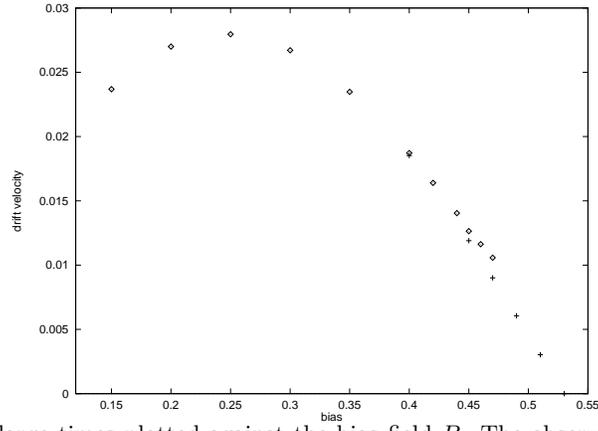,width=8cm,angle=270}
\caption{
The mean velocity  at large times plotted against the bias field
$B$. The observed values  in simulations at times of the order $10^9$ are
shown by diamonds. The extrapolated limiting values are shown by crosses.
Times are measured in units of diffusion attempts per walker, and distances
in units of the lattice constant.}
\end{center}
\end{figure}

\begin{figure}
\begin{center}
\leavevmode
\psfig{figure=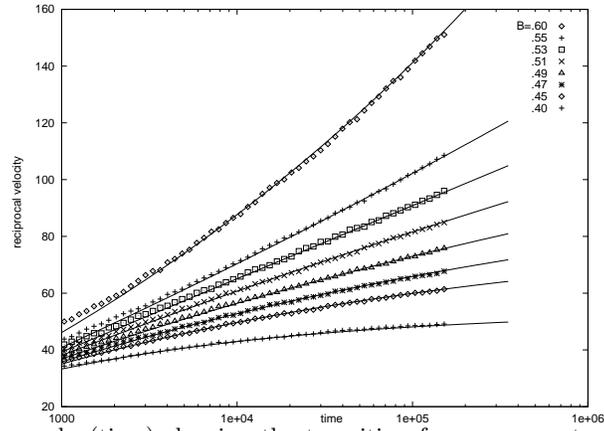,width=8cm,angle=270}
\caption{
Reciprocal velocity versus log(time) showing the transition from concave
to convex curvature at $B_c \simeq 0.53$ for intermediate times.}
\end{center}
\end{figure}

\begin{figure}
\begin{center}
\leavevmode
\psfig{figure=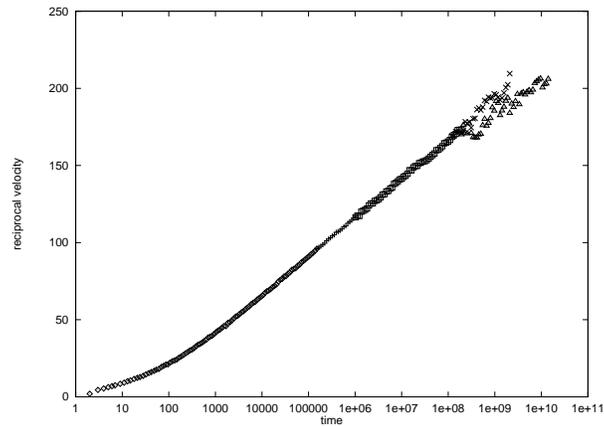,width=8cm,angle=270}
\caption{
Reciprocal velocity versus log(time) for $B=B_c$. Different symbols
show data averaged over different number $N$ of walkers. $N$ =  80,000
($\diamond$), 64,000 (+), 20000 (squares), 1024 ($\times$), 384 ($\triangle$).}
\end{center}
\end{figure}


\begin{thebibliography}{999}
\bibitem{reason} If Prob($\ell$) decreases with $\ell$ as $\ell^{-a}\;\exp(-A\ell)$,
this sum would vary as $\ell_{max}^{1-a}$. The connected part of the
two-point correlation function is known to decay as $R^{{1-d \over 2}}
\exp(-AR)$, for $AR >> 1$. To find the net probability that a site belongs
to a dead-end branch at a depth $\ell$, we have to integrate over $d-1$
transverse dimensions. This cancels the factor $R^{{{1-d \over 2}}}$, and
hence $a=0$ in all dimensions.
\end{thebibliography}
\end{document}